\documentstyle[aps,multicol]{revtex}
\input{psfig}

\def\be{\begin{equation}}
\def\ee{\end{equation}}
\def\beq{\begin{eqnarray}}
\def\eeq{\end{eqnarray}}

\begin{document}
\draft
\title{
DMRG Study of the Ground State
 at Higher Landau Levels\\
-- Stripes, Bubbles and the Wigner Crystal --
}
\author
{Daijiro {\sc Yoshioka} and Naokazu {\sc Shibata}}

\address{
Department of Basic Science, The University of Tokyo\\
3-8-1 Komaba, Meguro, Tokyo 153-8902, Japan
}

\date{June 4, 2001}

\maketitle
\begin{abstract}
Hartree-Fock theory predicted stripe or bubble phase in the third and higher Landau levels for two-dimensional electrons, and experimental evidences has been accumulated.
In this paper theoretical confirmation of the stripe phase and bubble phase in higher Landau level is given by means of the density matrix renormalization group (DMRG) method, which can give essentially exact ground state for electron systems with up to 18 electrons.
From the study of the pair correlation function, the stripe phase, bubble phase, and the Wigner crystal phase are identified, and phase diagram is obtained.
The reentrant integer quantum Hall state is identified as the bubble state.
The phase diagram of the fourth Landau level shows more diversity than the third level.
\end{abstract}

\begin{multicols}{2}

\narrowtext

%
%
\section{Introduction}
In the lowest Landau level the fractional quantum Hall effect is observed.
However, it has not been observed in the Landau levels with Landau index $N_{\rm L}$ larger than two.
This is because the interaction between electrons is not short-ranged enough to make Laughlin state realized.
In such higher Landau levels, Koulakov et al. proposed realization of the stripe and bubble phases based on the Hartree-Fock theory [1,2].
These phases belong to families of the charge density wave states proposed before the discovery of the fractional quantum Hall effect in the lowest Landau level [3,4].
So it is natural to anticipate such states once the FQH states are denied.

The evidence of the stripe phase has been obtained from experiments on ultra high mobility samples at low temperature as remarkable anisotropy in the longitudinal resistance when the higher Landau level is nearly half-filled [5,6].
Although no direct evidence of the bubble phase has been obtained, the reentrant IQHE around $\nu^* \simeq 1/4$ and 3/4 [7] suggests the realization of the bubble phase, where $\nu^*=\nu-[\nu]$ is the filling factor of the topmost partially filled Landau level.

Since, the Hartree-Fock theory may fail to find the true ground state as evidenced in the lowest Landau level, more reliable theory is desirable.
So exact diagonalization studies have been done, and the realization of the stripe phase and the bubble phase in the ground state was confirmed [8].
However, the system size treated by the exact diagonalization is quite small, and the filling factor studied was limited.
Because of these limitations, such study cannot determine the phase diagram of the present system.
In the present study we investigated the ground state by DMRG method [9].
We have investigated systems with up to 18 electrons, and determined the ground state at various filling factors by examining the pair correlation function.
Part of the investigation, where we clarified the phase diagram in the third lowest Landau level ($N_{\rm L}=2$) has already been published [10].
Here, we give new results for the fourth lowest landau level together with the previous results.
%
%
\section{Model and Method}
The Hamiltonian for electrons in Landau levels contains only the Coulomb interactions. After the projection onto the $N_{\rm L}$th Landau level, the Hamiltonian is written as
\begin{equation}
H=\sum_{i<j} \sum_{\bf q} e^{-q^2/2} \left[ L_{N_{\rm L}}(q^2/2) \right] ^2 V(q)
e^{i{\bf q} \cdot ({\bf R}_i-{\bf R}_j)} ,
\label{Coulomb}
\end{equation}
where ${\bf R}_i$ is the guiding center coordinate of the $i$th electron, which satisfy the commutation relation $[R_j^x,R_k^y] ={\rm i}\delta_{j,k}$. $L_{N_{\rm L}}(x)$ are the Laguerre polynomials, and $V(q) =2\pi e^2/q$ is the Fourier transform of the Coulomb interaction. The magnetic length $l$ is set to be 1.
We consider uniform positive background charge to cancel the component at $q=0$ in eq.~(\ref{Coulomb}), and neglect the electrons in fully occupied lower Landau levels.
Periodic boundary condition is imposed both in the $x$ and $y$ directions the period being $L_x$ and $L_y$, respectively.
The filling factor of the relevant Landau level is $\nu^*=2\pi N_{\rm e}/L_xL_y$, where $N_{\rm e}$ is the total number of electrons.
We used the DMRG method to obtain the ground state of the system with up to 18 electrons at various fillings with changing the aspect ratio $L_x/L_y$.

We first investigated the aspect ration dependence of the ground state energy at fixed filling factors.
Then pair correlation function $g(\bf r)$ in guiding center coordinates,
\begin{equation}
g({\bf r}) =\frac{L_xL_y}{N_{\rm e}(N_{\rm e}-1)} \langle \sum_{i \ne j}
\delta({\bf r}+{\bf R_i}-{\bf R_j})\rangle
\label{gr}
\end{equation}
is calculated at local minima of the ground state energy.
For the most filling factors our system is large enough that it is rather easy to tell the ground state CDW pattern by looking at $g({\bf r})$.

%
%
\section{Results}
From the pair correlation function we found that the ground state at $N_{\rm L}=2$ and 3 is either the Wigner crystal, the bubble state, or the stripe state depending on the filling factor $\nu^*$ and the Landau level index $N_{\rm L}$, except for a small region of the filling factor around 0.28 in $N_{\rm L}=3$ Landau level.
The resultant phase diagram is shown in Fig.1 and Fig.2.
For $N_{\rm L}=2$ Landau level there are three phases.
The boundary between the Wigner crystal state and the two-electron bubble phase lies between $\nu^*=2/9$ and $1/4$, and that between the two-electron bubble and the stripe phase lies between $\nu^*=4/11$ and $2/5$.
Inspection of the ${\bf r}$ dependence of $g({\bf r})$ shows that these boundaries are the first order phase transition points.
For $N_{\rm L}=3$ Landau level there are at least four phases.
Here three-electron bubble phase, where cluster of tree electrons form a kind of the Wigner crystal, becomes possible.
The ground state between $\nu^*=1/4$ and $0.3$ has not been identified.
We will discuss this state later.

\begin{figure}[b]
\psfig{figure=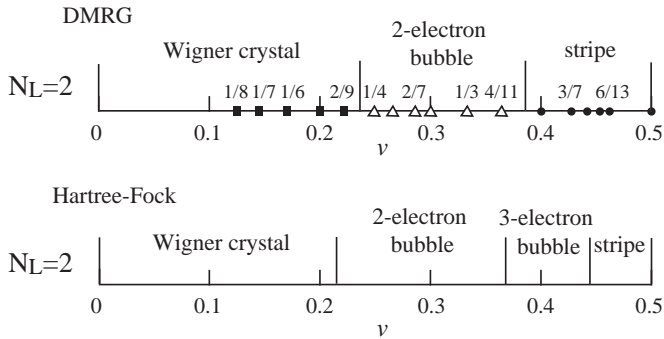,height=5cm}
\caption{The phase diagram for the third lowest Landau level ($N_{\rm L}=2$) obtained by DMRG and Hartree-Fock methods.}
\label{fig:1}
\end{figure}
\begin{figure}
\psfig{figure=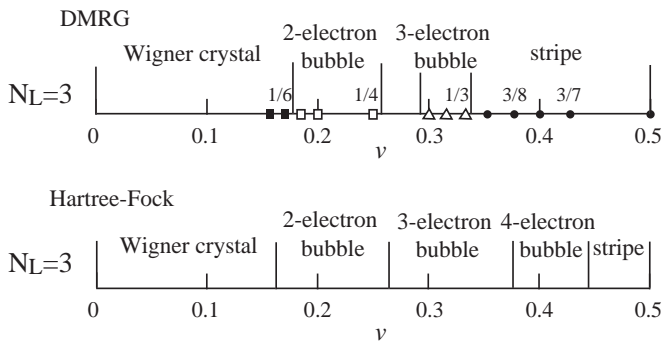,height=5cm}
\caption{The phase diagram for the third lowest Landau level ($N_{\rm L}=3$) obtained by DMRG and Hartree-Fock methods.}
\label{fig:2}
\end{figure}

We also solved the Hamiltonian by the Hartree-Fock theory [4].
In this method the filling factor is also limited to simple rational fractions.
However, for the bubble phase the order parameters obtained by the this method almost coincide with those proposed by Fogler and Koulakov [2], just the same as the case of the Wigner crystal state [4].
Then it is possible to obtain the Hartree-Fock ground state as a continuous function of the filling factor, and transition point between different bubble states can be determined.
The stripe phase has weak modulation along the stripes, which opens the gap at the Fermi level and lowers the energy.
The phase diagram by this method is also shown in Figs.1 and 2.
The phases are separated by the first order phase transition similarly to the DMRG case.
The phase diagrams from these two methods are similar except that the correct phase diagram by the DMRG shows wider region for the stripe phase.
The three-electron bubble phase at $N_{\rm L}=2$ and the four electron bubble phase at $N_{\rm L}=3$ are swallowed by the stripe phase.

%
%
\section{Discussion}
In this paper we have determined the phase diagram of electrons in the third and fourth lowest Landau levels by the DMRG method.
The slight difference in the phase diagrams for $N_{\rm L}=2$ and $3$ comes from the range of the Coulomb interaction projected to each Landau level, namely the Laguerre polynomial in eq.(\ref{Coulomb}) is responsible for the difference.
In actual systems there are several factors which affect the form of the Coulomb potential.
The first is the spread of the wave function in the direction perpendicular to the two-dimensional plane.
This spread acts mostly to reduce the short-range part of the Coulomb interaction.
Since this part is already reduced by the Laguerre factor in the higher Landau levels, we do not expect this will cause qualitative change to our phase diagram.
Another factor is the screening by the electrons in the filled lower Landau levels.
This acts to reduce the long-range part of the interaction.
We did not considered this factor, since this effect is sample dependent.
Namely, it depends on the electron density as well as the filling factor.
Comparing our Hartree-Fock results with those by Shklovskii's group, where the screening effect is considered [1,2], we have concluded that the screening effect does not affect the qualitative feature of the phase diagram.

We think that the reentrant IQH state is the two-electron bubble phase, which is pinned by impurities.
Then our calculation predicts that the reentrant phase should be observed in $N_{\rm L}=3$ Landau level at lower filling factor.

Our calculation until now have not identified the ground state between $\nu^*=1/4$ and $0.3$ for $N_{\rm L}=3$ Landau level.
The $g({\bf r})$ looks like stripe phase with stronger modulation.
Actually the energy of stripe phase by the Hartree-Fock theory is quite close to those of the 2-electron and 3-electron bubble phases here.
Since the energy lowering by the quantum fluctuation is larger in stripe phase as the result around $\nu^* \simeq 1/2$ shows, reentrant stripe phase is a feasible candidate.
We hope that we can identify this phase in near future.

\acknowledgements
Part of the numerical calculation is performed in the ISSP, University of Tokyo.
D.Y. thanks Aspen Center for Physics, where part of the work was done.
The present work is supported by Grant-in-Aid No.~12640308 and No.~11740184 from MEXT, Japan.

\end{multicols}


\begin{references}
\bibitem{ref:1} A.A. Koulakov, M.M. Fogler, and B.I. Shklovskii,
Phys.\ Rev.\ Lett. $\bf 76$, 499 (1996), and M.M. Fogler, A.A. Koulakov, B.I. Shklovskii, Phys.\ Rev.\ B $\bf 54$, 1853 (1996).
\bibitem{ref:2} M.M. Fogler and A.A. Koulakov, Phys.\ Rev.\ B $\bf 55$, 9326 (1997).
\bibitem{ref:3} H. Fukuyama, P.M. Platzman, and P.W. Anderson, Phys.\ Rev.\ B $\bf 19$, 5211 (1979).
\bibitem{ref:4} D. Yoshioka and H. Fukuyama, J.\ Phys.\ Soc.\ Jpn.\ {\bf 47}, 394 (1979), D. Yoshioka and P.A. Lee, Phys.\ Rev.\ B$\bf 27$, 4986 (1983).
\bibitem{ref:5} M.P. Lilly, K.B. Cooper, J.P. Eisenstein,
L.N. Pfeiffer, and K.W. West, Phys.\ Rev.\ Lett.\ $\bf 82$, 394 (1999).
\bibitem{ref:6} R.R. Du, D.C. Tsui, H.L. Stormer, L.N. Pfeiffer,
K. W. Baldwin, and K. W. West, Solid State Commun.\ $\bf 109$, 389 (1999).
\bibitem{ref:7} K.B. Cooper, M.P. Lilly, J.P. Eisenstein,
L.N. Pfeiffer, and K.W. West, Phys.\ Rev.\ B $\bf 60$, R11285 (1999).
\bibitem{ref:8} E.H. Rezayi, F.D.M. Haldane, and K. Yang,
Phys.\ Rev.\ Lett.\ $\bf 83$, 1219 (1999),
F.D.M. Haldane, E.H. Rezayi, and K. Yang, Phys.\ Rev.\ Lett.\ $\bf 85$, 5396 (2000).
\bibitem{ref:9} S.R. White, Phys.\ Rev.\ Lett.\ $\bf 69$ (1992) 2863;
Phys. Rev. B {\bf 48} (1993) 10345.
\bibitem{ref:10} N. Shibata and D. Yoshioka, Phys.\ Rev.\ Lett.\ {\bf 86} No.25 (2001).
\end{references}
\end{document}